\documentclass[11pt,letterpaper]{article}
\usepackage{color,ifpdf,latexsym,graphicx,url,amsmath,subfigure,float}

\usepackage{hyperref}

\advance \topmargin by -\headheight
\advance \topmargin by -\headsep
\evensidemargin \oddsidemargin
{\makeatletter \gdef\fps@figure{!htbp}}

\let\realbfseries=\bfseries
\def\bfseries{\realbfseries\boldmath}

\newtheorem{theorem}{Theorem}

\newtheorem{problem}[theorem]{Problem}

{\makeatletter
 \gdef\xxxmark{%
   \expandafter\ifx\csname @mpargs\endcsname\relax % in minipage?
     \expandafter\ifx\csname @captype\endcsname\relax % in figure/caption?
       \marginpar{xxx}% not in a caption or minipage, can use marginpar
     \else
       xxx % notice trailing space
     \fi
   \else
     xxx % notice trailing space
   \fi}
 \gdef\xxx{\@ifnextchar[\xxx@lab\xxx@nolab}
 \long\gdef\xxx@lab[#1]#2{\textbf{[\xxxmark #2 ---{\sc #1}]}}
 \long\gdef\xxx@nolab#1{\textbf{[\xxxmark #1]}}
}

\begin{document}

\title{A simple proof that the $(n^2-1)$-puzzle is hard}
\author{
Erik D. Demaine%
    \thanks{MIT Computer Science and Artificial Intelligence Laboratory,
      32 Vassar St., Cambridge, MA 02139, USA,
      \protect\url{edemaine@mit.edu}}
\and
  Mikhail Rudoy%
  \thanks{Google Inc, \protect\url{mrudoy@gmail.com}. Work completed at MIT Computer Science and Artificial Intelligence Laboratory.}
}
\date{}

\maketitle

\begin{abstract}
The 15 puzzle is a classic reconfiguration puzzle with fifteen uniquely labeled unit squares within a $4 \times 4$ board in which the goal is to slide the squares (without ever overlapping) into a target configuration. By generalizing the puzzle to an $n \times n$ board with $n^2-1$ squares, we can study the computational complexity of problems related to the puzzle; in particular, we consider the problem of determining whether a given end configuration can be reached from a given start configuration via at most a given number of moves. This problem was shown NP-complete in \cite{ratner}. We provide an alternative simpler proof of this fact by reduction from the rectilinear Steiner tree problem.
\end{abstract}

\section{Introduction}

The 15 puzzle is a classic puzzle consisting of fifteen sliding $1\times 1$ squares labeled with the numbers $1$ through $15$ within a $4 \times 4$ board. The remaining spot is taken up by a hole, whose presence allows the other squares to slide. The goal of the puzzle is to slide the squares in such a way that they end up arranged in order, while avoiding collisions.

We can naturally generalize the 15 puzzle to the $(n^2 - 1)$-puzzle, in which $n^2 - 1$ sliding squares and one hole fill an $n \times n$ board of cells. Given a pair of configurations (injective mappings from squares to cells), whether one can be reached from the other by any sequence of moves can be decided in polynomial time (see for example \cite{archer}). But what if we want to minimize the number of moves in the sequence? Adding a bound on the number of moves used to transition between the configurations, we obtain the following more interesting decision problem:

\begin{problem}
The \emph{$(n^2 - 1)$-puzzle problem} asks, for a given pair of puzzle configurations $s$ and $t$ and a given number $k$, whether it is possible to convert configuration $s$ into configuration $t$ using at most $k$ moves.
\end{problem}

This problem was shown NP-complete in \cite{ratner} by reduction from a rather complicated variant of SAT introduced in that paper. The purpose of this paper is to provide a simpler proof of the same fact. In particular, we provide a reduction from the rectilinear Steiner tree problem to the $(n^2 - 1)$-puzzle problem.

\begin{problem}
The \emph{rectilinear Steiner tree problem} asks, for a given number $l$ and a given set of points with positive integer coordinates in the plane, whether it is possible to find a tree in the plane whose total length does not exceed $l$, whose edges are all rectilinear (i.e., horizontal or vertical), and which passes through each of the given points.
\end{problem}

The rectilinear Steiner tree problem was shown strongly NP-hard in \cite{steiner}, so a reduction from this problem to the $(n^2 - 1)$-puzzle problem is sufficient for our desired result.

\section{Reduction}

Suppose we are given a rectilinear Steiner tree instance consisting of a set of points $P = (p_1, p_2, \ldots, p_{|P|})$ and a value $l$. We wish to systematically convert the pair $(P, l)$ into a $(n^2 - 1)$-puzzle instance consisting of a pair of puzzle configurations $s$ and $t$ and a value $k$.

Let $c = 18 \cdot |P|$, and let $m$ be the maximum coordinate of any point in $P$. Then the puzzle configurations $s$ and $t$ that we produce will be $n \times n$ puzzles with $n = (m+1) \cdot c$. 

Now we define a correspondence between the points in $P$ and some of the cells of the puzzle board. In particular, we correspond point $(x, y)$ with the puzzle board cell at location $(cx, cy)$. Notice that this location is within the board because, by design, $cx$ and $cy$ are both bounded above by $m \cdot c$.

\begin{figure}[!hbt]
\centering
\includegraphics[scale=.25]{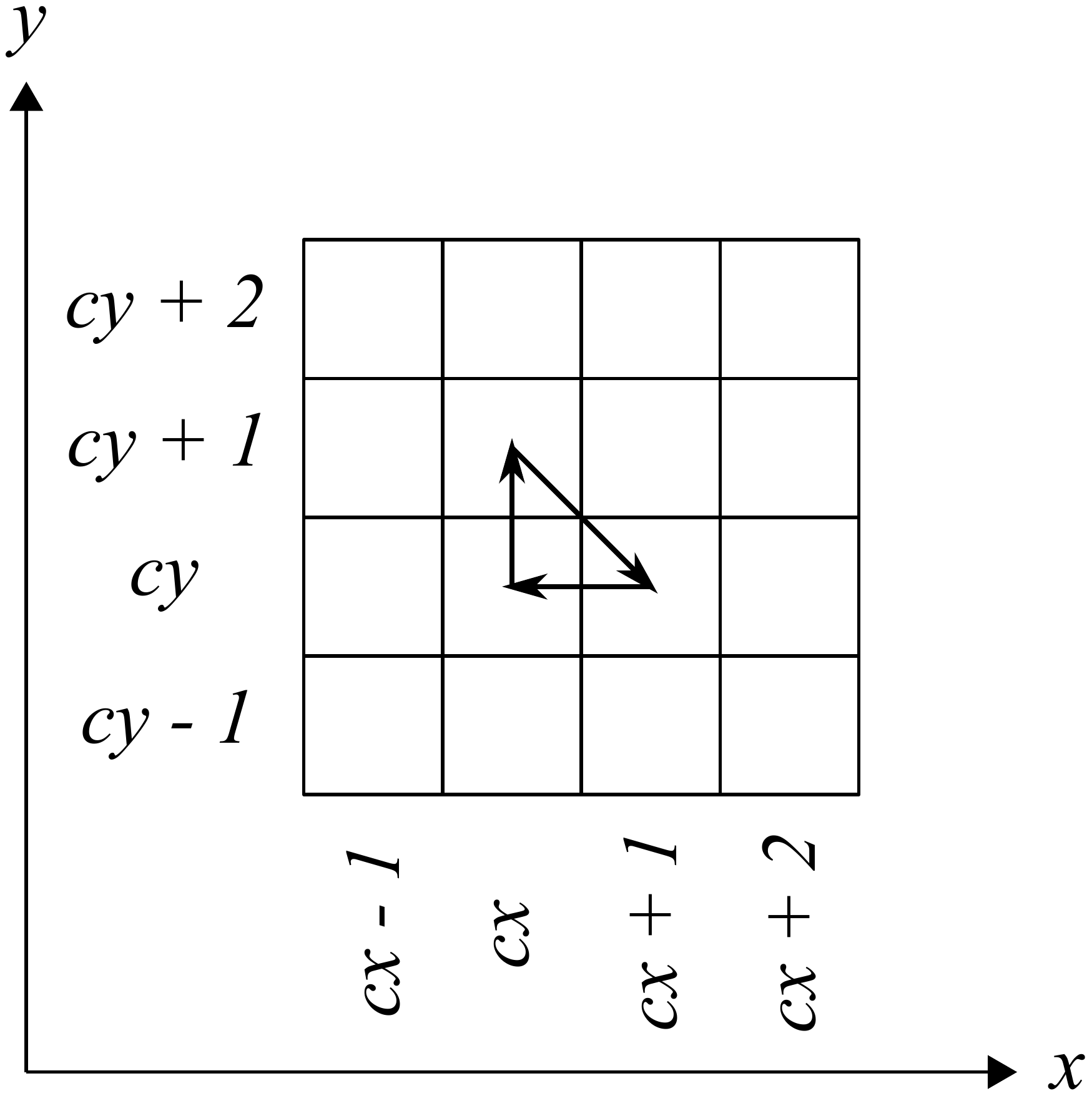}
\caption{
Configuration $t$ is the same as configuration $s$, except near cells with coordinates $(cx, cy) \in P$. For each such pair in $P$, three squares in $t$ are cycled relative to their positions in $s$ as shown here.
\label{fig:cycle}}
\end{figure}

With this correspondence in place, we can define configurations $s$ and $t$. The hole in $s$ is placed in the location corresponding to point $p_1$. We number every other cell in $s$ arbitrarily. To construct $t$, we modify configuration $s$. For each point $p \in P \setminus \{p_1\}$, permute the three squares in the board locations directly above, directly to the right, and at the puzzle board cell corresponding to $p$ in a clockwise cycle. This is shown in Figure~\ref{fig:cycle}. The result of doing this operation to $s$ for every $p \in P \setminus \{p_1\}$ is the puzzle configuration $t$. 

Finally, define the move bound $k = (2l+1)\cdot c$ to conclude the reduction.

\section{Polynomial time}

Notice that the board size is polynomial in the maximum coordinate value $m$ and the number of points $|P|$. Because the rectilinear Steiner tree problem is strongly NP-hard, we can assume that both $m$ and $|P|$---and therefore also the board size $n$---are polynomial in the size of the input.

Since the contents of every board cell in $s$ or $t$ can be computed easily, and since the number of board cells is polynomial in the size of the input, both $s$ and $t$ can be computed from $(P, l)$ in polynomial time. Furthermore, $k$ can be computed easily in polynomial time, so the whole reduction runs in polynomial time.

\section{Rectilinear Steiner tree instance solution $\to$ $(n^2 - 1)$-puzzle instance solution}

Suppose that there is a solution to the input rectilinear Steiner tree instance. Hanan's Lemma \cite[Theorem 4]{Hanan-1966} states that there exists an optimal-length rectilinear Steiner tree through the points in $P$ whose edges all follow the grid formed by a horizontal and a vertical line
through each point in $P$. In other words, there exists a rectilinear Steiner tree $T$ with unit-length edges over at most $l+1$ vertices with coordinates in $\{1, \ldots, m\}$ which touches each point in $P$.

\begin{figure}[!hbt]
\centering
\includegraphics[scale=.5]{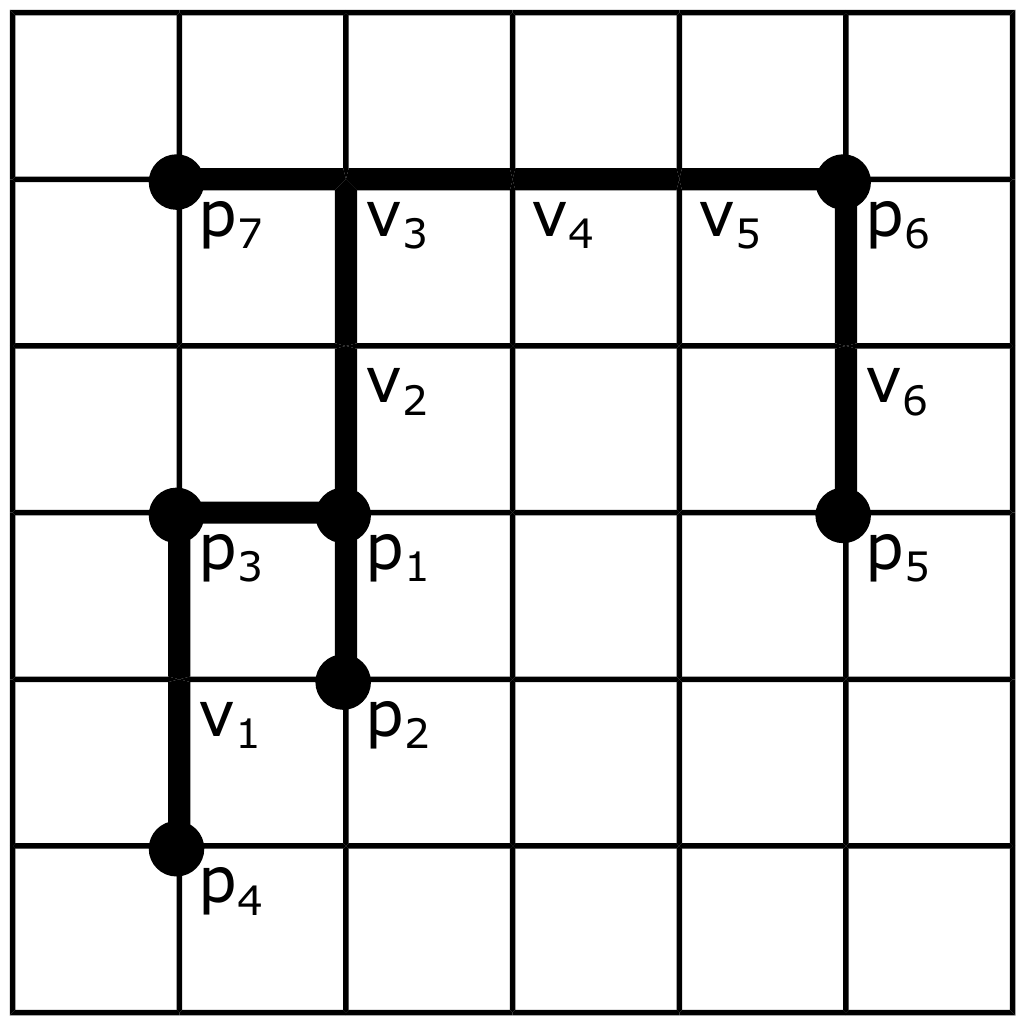}
\caption{
An example rectilinear Steiner tree for the points $p_1, p_2, \ldots, p_7$ which also has additional integer coordinate vertices $v_1, \ldots, v_6$. Every edge of this tree has length 1. 
\label{fig:traversal_example}}
\end{figure}

We will use this tree to construct a sequence $S'$ of moves solving the $(n^2 - 1)$-puzzle instance. We begin by building a simpler sequence of moves $S$. Construct an Euler tour of tree $T$ rooted at $p_1$ (i.e., a traversal of the edges of $T$ starting and ending at $p_1$ which passes through each edge exactly once in each direction). For example, one possible traversal for the tree in Figure~\ref{fig:traversal_example} is
$$p_1, p_3, v_1, p_4, v_1, p_3, p_1, p_2, p_1, v_2, v_3, p_7, v_3, v_4, v_5, p_6, v_6, p_5, v_6, p_6, v_5, v_4, v_3, v_2, p_1.$$
For each edge in the Euler tour, add $c$ moves to $S$ which move the hole in correspondence to the direction of the traversal: if the traversal moves from vertex $(x, y)$ to vertex $(x \pm 1, y)$, then move the hole directly right/left from cell $(cx, cy)$ to cell $(c(x \pm 1), cy)$, and if the traversal moves from vertex $(x, y)$ to vertex $(x, y \pm 1)$, then move the hole directly up/down from cell $(cx, cy)$ to cell $(cx, c(y \pm 1))$. Notice that the hole starts at cell $(cx, cy)$ where $(x, y) = p_1$ and that the traversal starts at $p_1$. Thus this definition of the moves in $S$ is well defined.

What is the effect of move sequence $S$? Just as the traversal passes each edge twice (once in each direction), the effect of the move sequence is that the hole passes each horizontal/vertical path of cells twice (once in each direction). The second time the hole passes some path of cells it exactly undoes the effect of the first pass. Thus the overall effect of $S$ is the identity transformation. Below, we modify $S$ into the $(n^2-1)$-puzzle instance solution $S'$ by adding moves.

Consider any point $p_i = (x, y) \in P \setminus \{p_1\}$. For any such point, we can consider the $4 \times 4$ window of cells with $x$ coordinate in the set $\{cx-1,cx,cx+1,cx+2\}$ and $y$ coordinate in the set $\{cy-1,cy,cy+1,cy+2\}$. Call this window $w_i$. Under move sequence $S$, the hole visits window $w_i$ but does not end inside it. Therefore there is some move in $S$ which moves the hole out of window $w_i$ for the final time. We modify $S$ by adding an extra sequence of moves $E_i$ immediately before this move. The overall effect of move sequence $E_i$ will be to cycle the squares in cells $(cx, cy)$, $(cx+1, cy)$, and $(cx, cy+1)$ with $(x,y) = p_i$ and to not move any other squares (or the hole). Thus, when resuming $S$ after $E_i$, the behavior of $S$ will be the same as before (except that those $3$ squares will have been permute). Because the effect of $S$ on its own is the identity transformation, the overall effect of the modified move sequence $S'$ is to cycle the squares in cells $(cx, cy)$, $(cx+1, cy)$, and $(cx, cy+1)$ for each $(x,y) = p_i \in P \setminus \{p_1\}$. In other words, $S'$ will accomplish exactly the transformation needed to transform configuration $s$ into configuration $t$.

\begin{figure}
  \centering
  \begin{minipage}{0.35\linewidth}
    \centering
    \includegraphics[scale=.25]{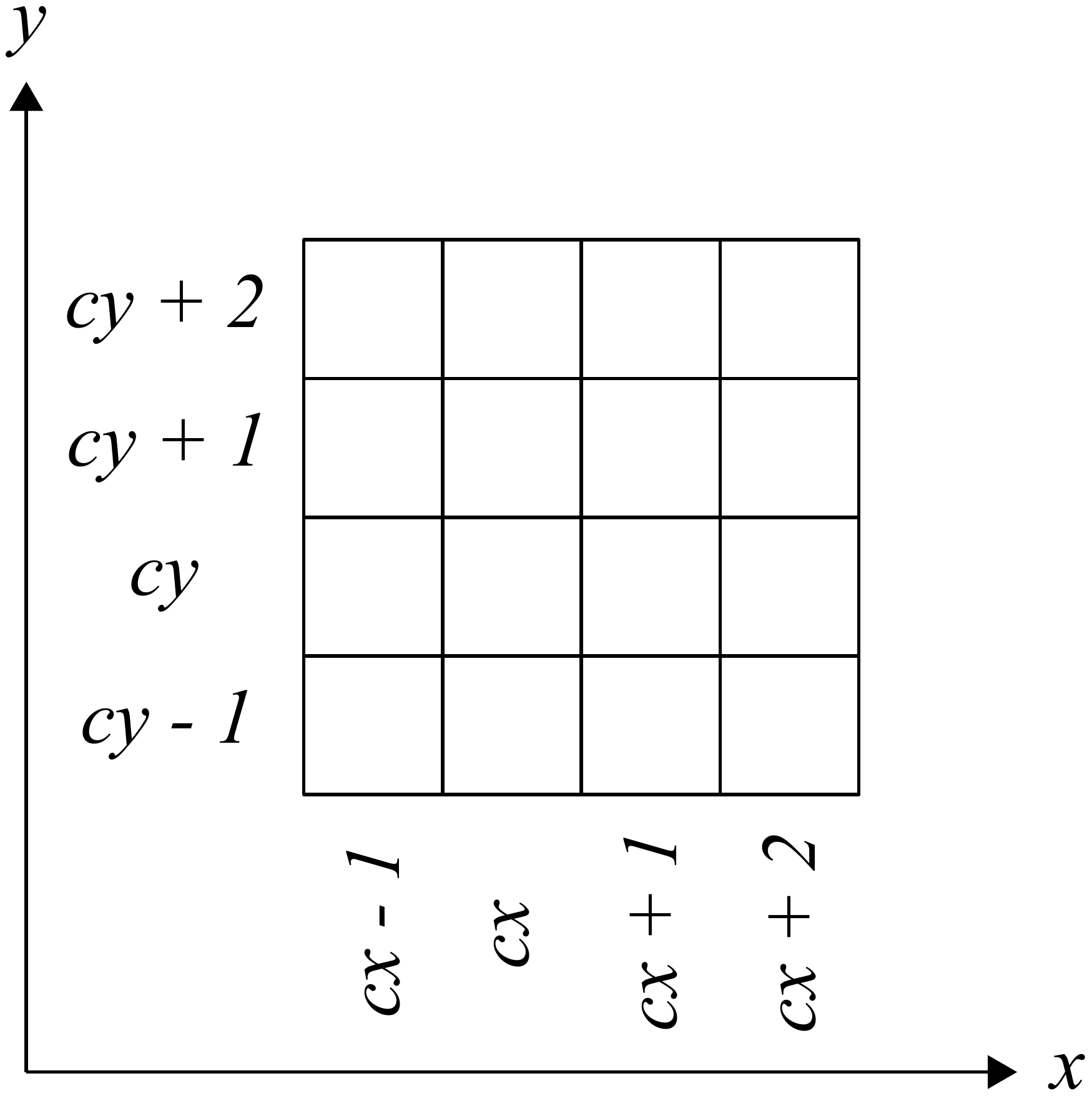}
  \end{minipage}\hfill
  \begin{minipage}{0.5\linewidth}
    \centering
    \includegraphics[scale=.25]{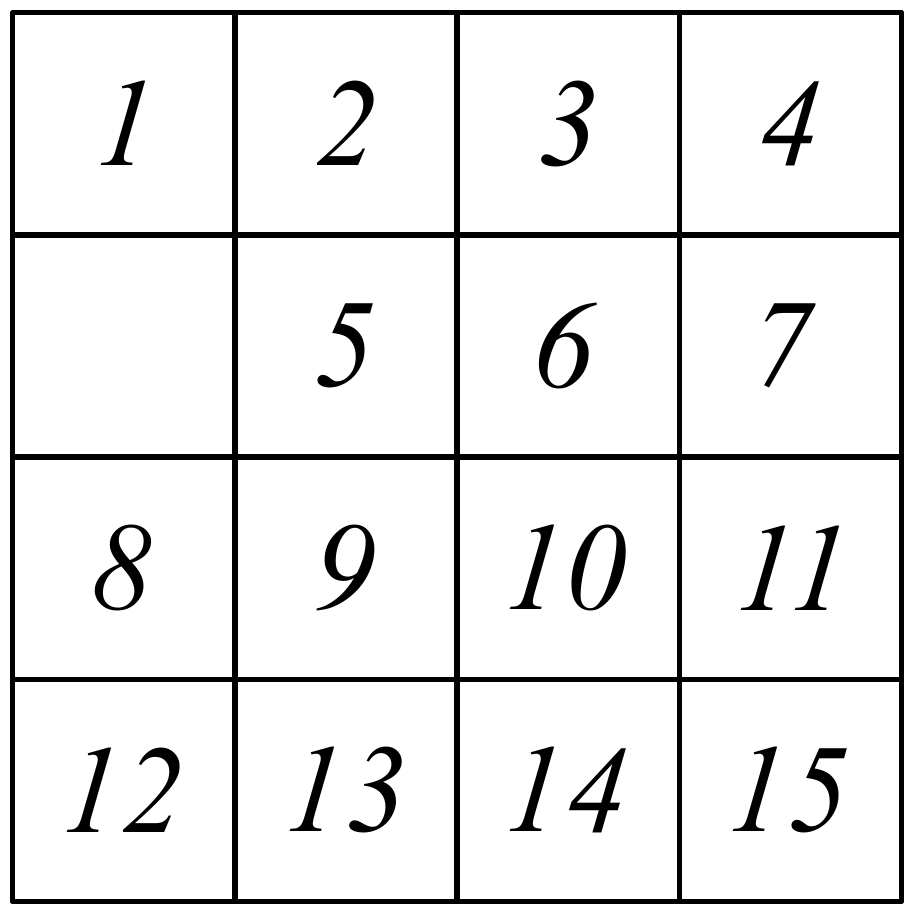}\hfill
    \includegraphics[scale=.25]{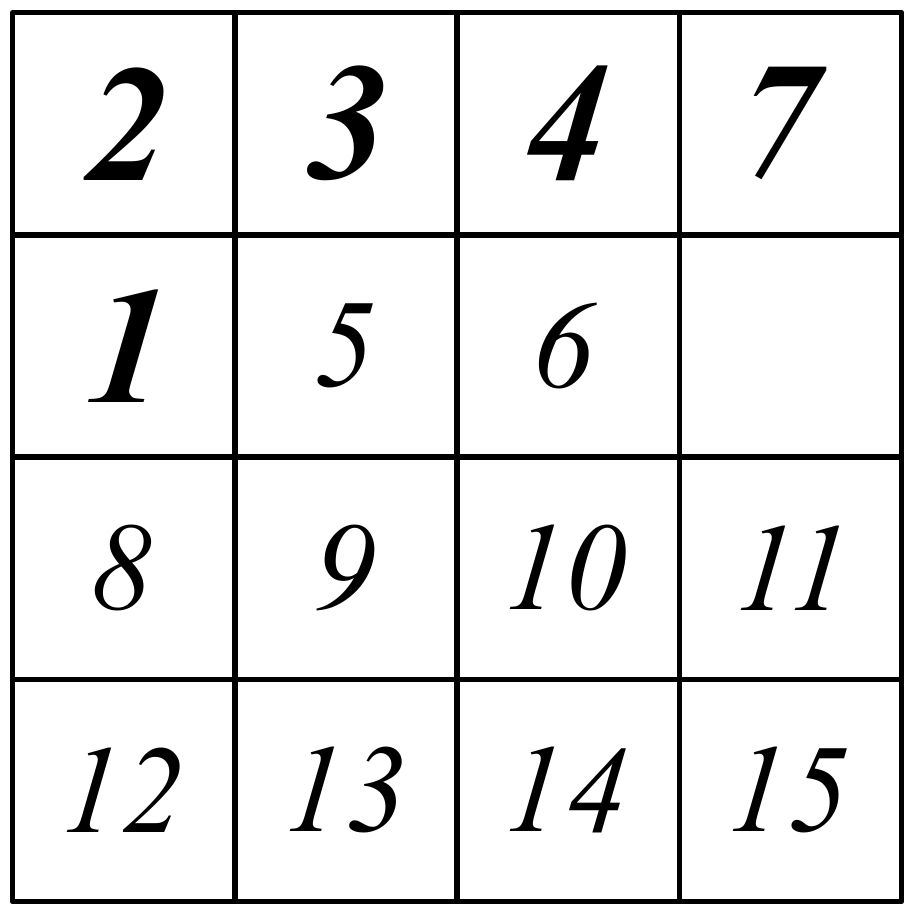}\hfill
    \includegraphics[scale=.25]{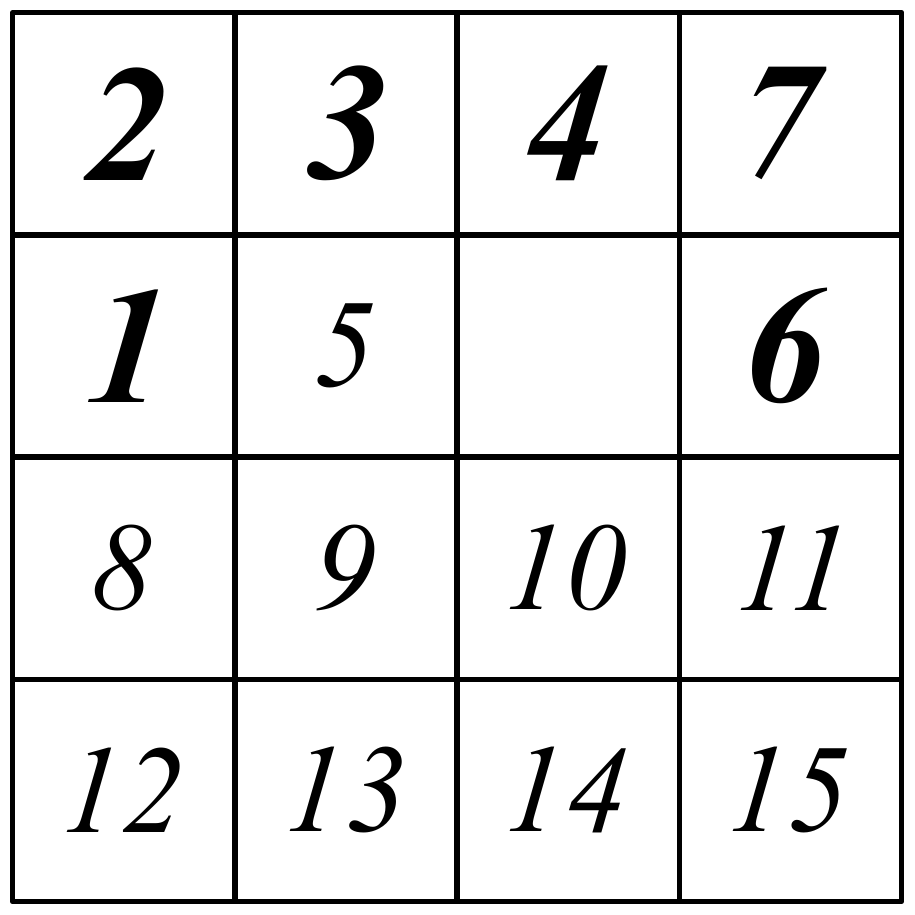}\vspace{0.5cm}
    \includegraphics[scale=.25]{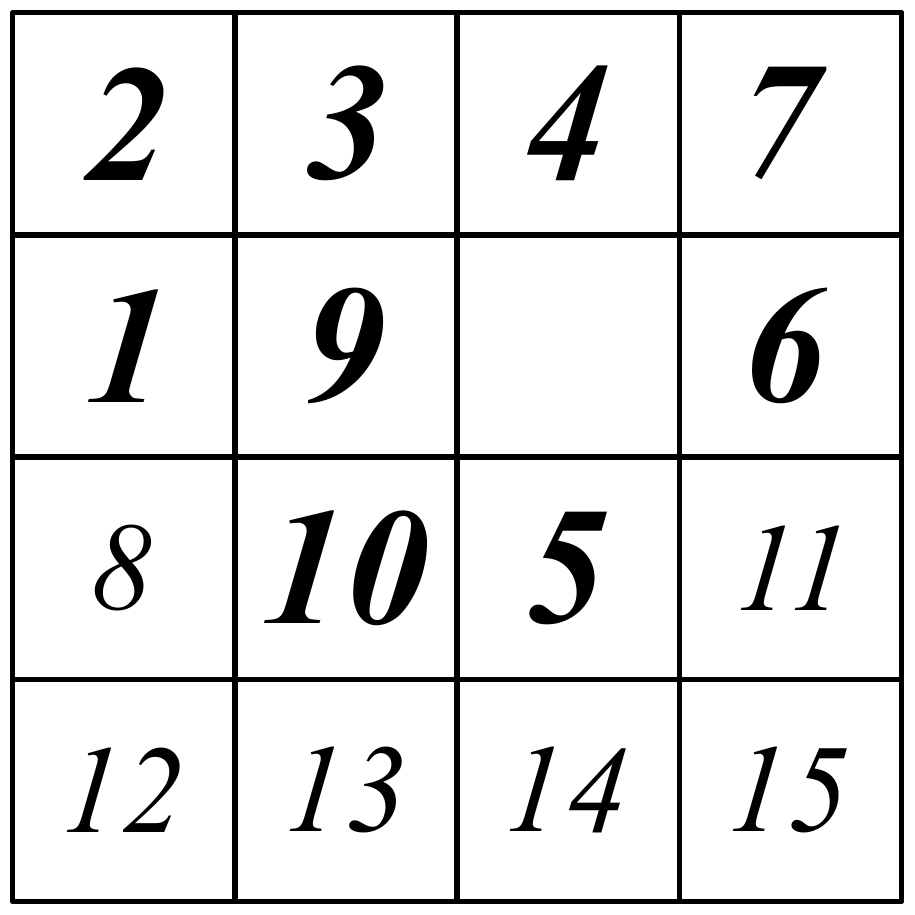}\hfill
    \includegraphics[scale=.25]{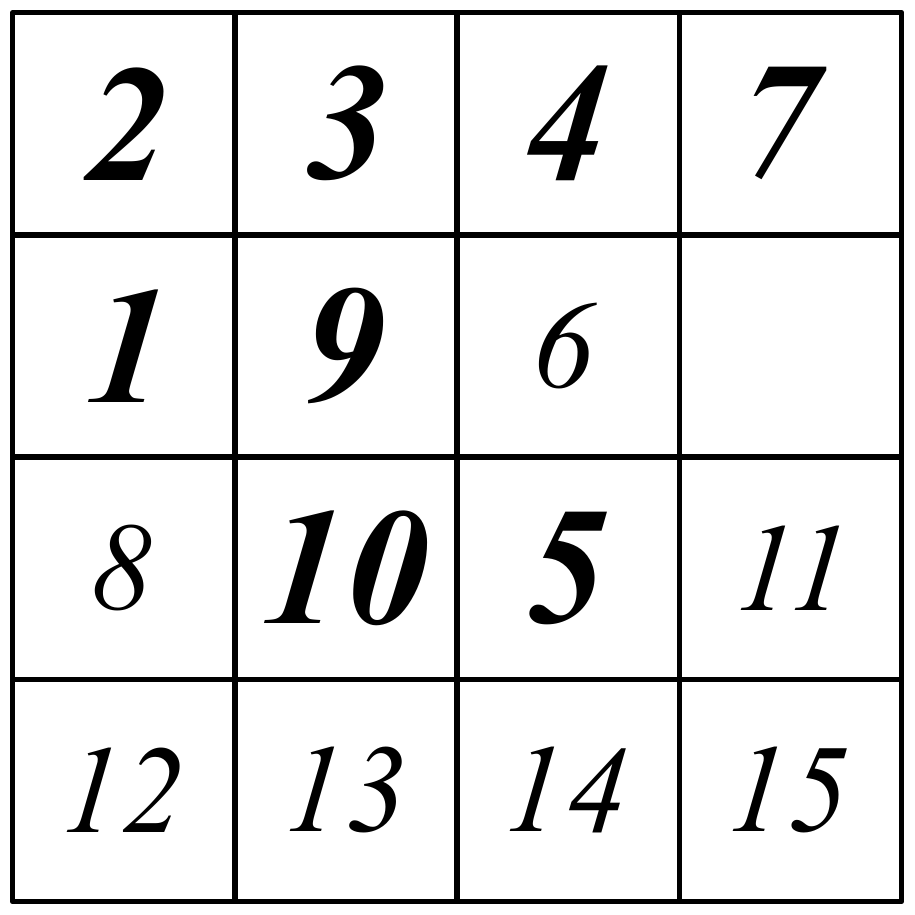}\hfill
    \includegraphics[scale=.25]{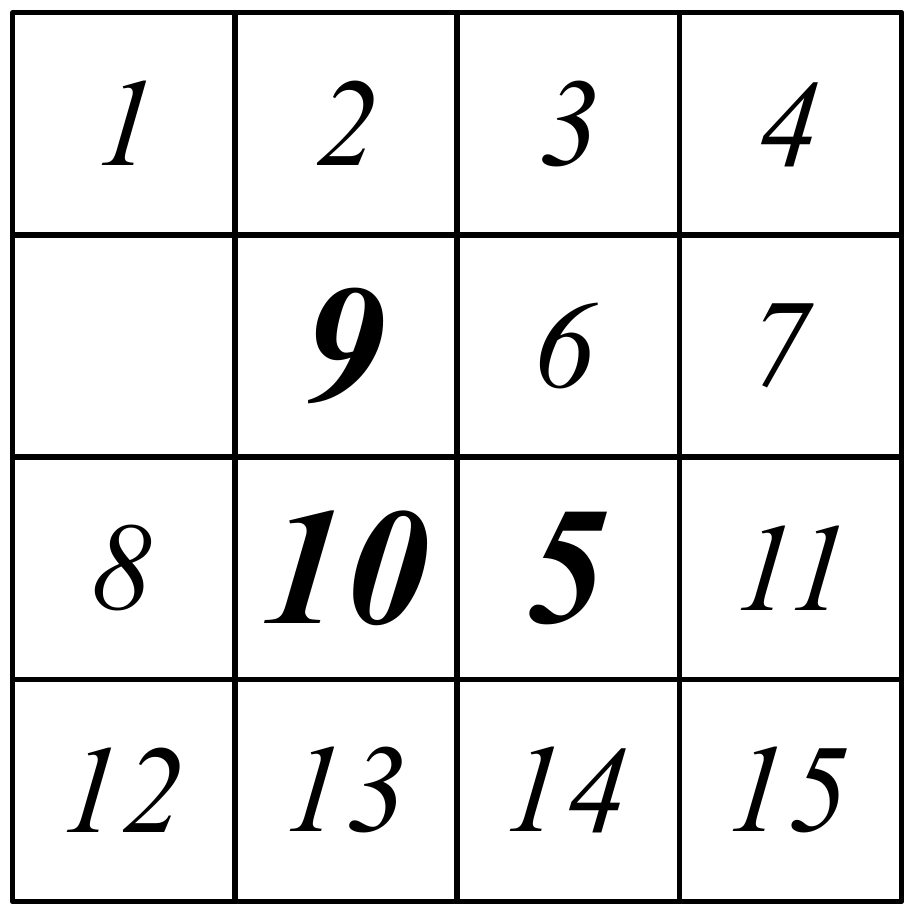}
  \end{minipage}
  \caption{An example move sequence $E_i$ within $4 \times 4$ window $w_i$ (left) using numbers $1$ through $15$ to represent the labels of the squares in the window. In each board (right), the bold labels indicate squares that are displaced from their original position in the first board. In the first board, we show an example of what the window might look like immediately before $E_i$ in the case that the hole starts in cell $(cx-1, cy+1)$. In the second board, we show the window after the hole has moved to cell $(cx+2, cy+1)$ along the boundary of the board. In the third board, we show the window after one more step (with the hole moving to $(cx+1, cy+1)$). The fourth board shows the window after the left, down, right, up movement of the hole. The fifth board shows the window after the hole moves back to cell $(cx+2, cy+1)$. The final board shows the window after the hole has moved back to its initial position. This is the final state of the window after all of $E_i$.}
  \label{fig:E_i_example}
\end{figure}

Immediately before $E_i$ is the last moment when the hole is in window $w_i$. Thus, the hole must be somewhere among the 12 boundary cells of the window; see Figure~\ref{fig:E_i_example} for an example. We begin $E_i$ by moving the hole along this 12 cell boundary until it is in cell $(cx+2, cy+1)$. This takes at most 6 moves. Next, we move the hole left one cell into position $(cx+1, cy+1)$. 
%Note that, so far, $E_i$ has not affected the squares in cells $(cx, cy)$, $(cx+1, cy)$, and $(cx, cy+1)$. 
%Furthermore, $S$ would not have affected these cells any more if we were to just continue $S$ instead of pausing for $E_i$, so since the end configuration resulting from $S$ is $s$, these cells currently contain the same squares that they contain in $s$. 
Next, the hole moves in a square pattern: left, down, right, and up. This puts the hole back in the exact same position, but cycles the three squares in the positions it passes through. In particular, the squares in cells $(cx, cy)$, $(cx+1, cy)$, and $(cx, cy+1)$ are cycled clockwise. After that, we undo the initial part of $E_i$: the hole moves right to $(cx+2, cy+1)$ and then retraces its steps to wherever it started before $E_i$. 
Note that the only part of $E_i$ affecting the squares in cells $(cx, cy)$, $(cx+1, cy)$, and $(cx, cy+1)$ is the left-down-right-up motion of the hole. Thus, the overall effect of $E_i$ is to cycle the squares in cells $(cx, cy)$, $(cx+1, cy)$, and $(cx, cy+1)$ clockwise and to leave the other squares unaffected.
In the example of Figure~\ref{fig:E_i_example}, the desired behavior is to cycle the squares $5$, $9$, and $10$ (the squares in cells $(cx, cy)$, $(cx+1, cy)$, and $(cx, cy+1)$) clockwise, which is exactly what happens by the end of the example.

Because $E_i$ cycles the squares in cells $(cx, cy)$, $(cx+1, cy)$, and $(cx, cy+1)$ without affecting anything else, move sequence $S'$ converts configuration $s$ into configuration $t$ as argues above. Furthermore, the number of moves in $S'$ is equal to the number of moves in $S$ plus the number of moves in each $E_i$ for $i = 2, 3, \ldots, |P|$. The number of moves in $S$ is $2c$ times the number of edges in $T$. Thus the number of moves in $S$ is at most $2cl$. The number of moves in $E_i$ is at most $18$. Thus the number of moves in all of the $E_i$s together is at most $18(|P| - 1) < c$. In other words, the total number of moves in $S'$ is less than $2cl+c = (2l+1) \cdot c = k$. Thus, $S'$ solves the $n^2-1$ puzzle instance.

\section{$(n^2 - 1)$-puzzle instance solution $\to$ rectilinear Steiner tree instance solution}

Suppose there exists a solution to the $(n^2 - 1)$-puzzle instance. Consider a solution $S$ using the minimum possible number of moves (where this quantity is less than or equal to $k$). Let $R$ be the set of cells visited by the hole over the course of solution $S$. 

Define a cell of the board to be \emph{$(s,t)$-similar} if it contains a square in both $s$ and $t$ and if this square is the same square in both configurations. Notice that every non-$(s,t)$-similar cell $a$ is in $R$. This is because there are only two ways for cell $a$ to be non-$(s,t)$-similar: either cell $a$ contains the hole in $s$ or $t$ or cell $a$ contains two different squares in $s$ and $t$. If cell $a$ contains a hole in $s$ or $t$, then the hole visits that cell over the course of solution $S$, and so by definition, $a \in R$. If cell $a$ contains two different squares in $s$ and $t$ then the hole must pass through $a$ in order to move the square placed there in $s$. In this case too we see that $a \in R$.

Let $V$ be the set of coordinate pairs $(x,y)$ of cells in $R$ and let $G$ be the grid graph induced by vertex set $V$ (with edges between vertex pairs at unit distance). Because the region visited by the hole over the course of solution $S$ (the region of $R$) must be a connected region, we know that $G$ must be a connected graph. Consider any spanning tree $T$ of this graph. The spanning tree passes through all $|R|$ vertices of $G$, and has $|R| - 1$ edges. Each edge of this tree has length $1$, so we see that $T$ is a tree in the plane of total length $|R| - 1$ which passes through every point $(x,y)$ such that $(x,y)$ are the coordinates of a cell in $R$. Notice that the cell in position $(cx, cy)$ where $(x, y) \in P$ is non-$(s,t)$-similar, and therefore in $R$. In other words, the tree $T$ of total length $|R| - 1$ passes through every point of the form $(cx, cy)$ where $(x,y) \in P$.

Scaling $T$ down by a factor of $c$, we obtain a new tree $T'$ which passes through every point of the form $(x, y)$ where $(x,y) \in P$, whose edges are all vertical or horizontal, and whose total length is strictly less than $\frac{|R|}{c}$. In other words, we have a rectilinear Steiner tree passing through each point of $P$ of total length less than $\frac{|R|}{c}$. The optimal rectilinear Steiner tree through the points in $P$ has integer length, so to solve the rectilinear Steiner tree instance (using length at most $l$) it suffices to show that $\frac{|R|}{c} \le l+1$.

In order to show that $\frac{|R|}{c} \le l+1$, we need to bound $|R|$, which we will do by bounding the number of $(s,t)$-similar and non-$(s,t)$-similar cells in $R$. Let $R' \subseteq R$ be the set of $(s,t)$-similar cells in $R$, so $R \setminus R'$ is the set of non-$(s,t)$-similar cells (in $R$). 

First, we compute an upper bound on $|R \setminus R'|$. Every cell $a \in R \setminus R'$ is by definition non-$(s,t)$-similar. Thus, $|R \setminus R'|$ is at most the total number of non-$(s,t)$-similar cells in the whole board. During the construction of $t$ from $s$, each point in $P \setminus \{p_1\}$ contributes three non-$(s,t)$-similar cells (namely, the three cells whose squares are permuted). In addition, the cell containing the hole in both $s$ and $t$ is also non-$(s,t)$-similar. There are no other non-$(s,t)$-similar cells in the whole board. Thus, there are exactly $3(|P| - 1) + 1 = 3|P| - 2$ cells that are non-$(s,t)$-similar. As a result, $|R \setminus R'| \le 3|P| - 2 \le \frac{c}{2}$, because $c = 18|P|$.

Next we compute an upper bound for $|R'|$. Consider any cell $a \in R'$. Because $a \in R' \subseteq R$, the hole must enter this cell at least once over the course of solution $S$. We wish to show that the hole must actually enter the cell at least twice. Because there are only $k$ moves, we will conclude that $|R'| \le \frac{k}{2}$. 

Suppose for the sake of contradiction that the hole enters cell $a$ exactly once. Then at some point in $S$, the hole enters cell $a$ from a neighboring cell $a_{\mathrm{before}}$, and then exits into cell $a_{\mathrm{after}}$. If $a_{\mathrm{before}} = a_{\mathrm{after}}$, then omitting these two moves leaves the overall behavior of the solution the same. This cannot be the case because $S$ is a minimal move solution. Thus $a_{\mathrm{before}} \ne a_{\mathrm{after}}$. When the hole moves from $a_{\mathrm{before}}$ into $a$, the square initially in $a$ is moved to $a_{\mathrm{before}}$. Then in the next move, the hole moves from $a$ to $a_{\mathrm{after}}$ and the square in $a_{\mathrm{after}}$ moves into $a$. Because $a_{\mathrm{before}} \ne a_{\mathrm{after}}$, this means that the square which moves into $a$ due to these two moves is not the square that started there. But since the hole never again enters $a$, the square now in $a$ never exits it. This means that the square in cell $a$ is different in the initial configuration $s$ and the final configuration $t$. That, however, contradicts the fact that $a$ must be $(s,t)$-similar. As desired, this means that the hole must enter cell $a$ at least twice.

Because the hole enters every $a \in R'$ at least twice and there are only $k$ moves, we conclude that $|R'| \le \frac{k}{2}$. 

We showed above that $|R \setminus R'| \le \frac{c}{2}$ and $|R'| \le \frac{k}{2}$. Putting this together, we see that $|R| = |R'| + |R \setminus R'| \le \frac{k}{2} + \frac{c}{2} = \frac{(2l+1) \cdot c}{2} + \frac{c}{2} = (l+1) \cdot c$. As a result, we have that $\frac{|R|}{c} \le l+1$, which is exactly what we wished to prove.

\section{Conclusion}

We have shown that the given reduction is answer preserving and polynomial-time computable. Therefore, the reduction from rectilinear Steiner tree is sufficient to prove that the $(n^2 - 1)$-puzzle problem is NP-hard.

A natural open question is to ask what number of moves is required to solve an $(n^2-1)$-puzzle in the worst case as a function of $n$. For $n = 4$, this number was shown to be $80$ in \cite{god_number}. This function is known to be at most $O(n^3)$ by the result from \cite{parberry} and can be shown to be $\Omega(n^2)$ by a simple potential argument.

\bibliography{puzzle}
\bibliographystyle{plain}

\end{document}